\documentclass[fleqn,usenatbib]{mnras}

\usepackage{graphicx}
\usepackage{amsmath}	

\usepackage{txfonts}

\usepackage[T1]{fontenc}

\DeclareRobustCommand{\VAN}[3]{#2}
\let\VANthebibliography\thebibliography
\def\thebibliography{\DeclareRobustCommand{\VAN}[3]{##3}\VANthebibliography}

\newcommand{\msun}{$M_\odot$}

\newcommand{\hii}{H\,{\sc ii}\rm}

\newcommand{\nii}{[N\,{\sc ii}]}

\newcommand{\oiii}{[O\,{\sc iii}]}
\newcommand{\oii}{[O\,{\sc ii}]}

\newcommand{\sii}{[S\,{\sc ii}]}

\newcommand{\op}{O$^{+}$}

\newcommand{\np}{N$^{+}$}

\newcommand{\eg}{{e.g.}}
\newcommand{\ie}{{i.e.}}

\newcommand{\te}{$\rm T_e$}

\newcommand{\lin}{$\,\lambda$}
\newcommand{\llin}{$\,\lambda\lambda$}

\newcommand{\rtf}{$r_{25}$}

\newcommand{\rtwothree}{R$_{23}$}
\newcommand{\vs}{vs.}

\renewcommand{\eg}{\mbox{e.g.}}
\renewcommand{\ie}{\mbox{i.e.}}
\newcommand{\fwhm}{\mbox{\small FWHM}}

\newcommand{\halpha}{H$\alpha$}
\newcommand{\hbeta}{H$\beta$}

\newcommand{\kms}{\mbox{km\,s$^{-1}$}}
\newcommand{\uflux}{\rm erg\,cm$^{-2}$\,s$^{-1}$}

\title[Mixing in NGC 6946]{SIGNALS on the mixing of oxygen and nitrogen in the spiral galaxy NGC~6946}

\author[F. Bresolin et al.]{Fabio Bresolin,$^{1}$\thanks{E-mail: bresolin@ifa.hawaii.edu}
	David Fern\'andez-Arenas,$^{2}$
	Laurie Rousseau-Nepton,$^{3,4,2}$
	Ray Garner III,$^{5,6}$\newauthor
	Almudena Zurita,$^{7,8}$ 
	Carmelle Robert,$^{9,10}$
	Laurent Drissen,$^{9,10}$
	Ren\'e Pierre Martin,$^{11}$\newauthor
	Philippe Amram,$^{12}$
	Salvador Duarte Puertas,$^{9,10,7,8}$
	Gabriel Savard,$^{9,10}$\newauthor
	S\'ebastien Vicens,$^{9,10}$
	Mykola Posternak$^{9,10}$\medskip
	\\
	$^{1}$Institute for Astronomy, University of Hawaii, 2680 Woodlawn Drive, Honolulu HI 96822, USA \\
	$^{2}$Canada-France-Hawaii Telescope, 65-1238 Mamalahoa Hwy, Kamuela HI 96743, USA\\
	$^{3}$Dunlap Institute of Astronomy and Astrophysics, University of Toronto, 50 St. George St., Toronto, ON, M5S 3H4, Canada\\
	$^{4}$Department of Astronomy \& Astrophysics, University of Toronto, 50 St. George St., Toronto, ON, M5S 3H4, Canada\\
	$^{5}$Department of Physics and Astronomy, Texas A\&M University, 578 University Dr., College Station, TX 77843, USA\\
	$^{6}$George P. and Cynthia W. Mitchell Institute for Fundamental Physics \& Astronomy, Texas A\&M University, 578 University Dr., College Station, TX 77843, USA\\
	$^{7}$Departamento de F\'isica Te\'orica y del Cosmos, Campus de Fuentenueva, Edificio Mecenas, Universidad de Granada, E-18071 Granada, Spain\\
	$^{8}$Instituto Carlos I de F\'isica Te\'orica y Computacional, Facultad de Ciencias, Universidad de Granada, E-18071 Granada, Spain\\
	$^{9}$D\'epartement de Physique, de G\'enie Physique et d’Optique, Universit\'e Laval, Canada\\
	$^{10}$Centre de Recherche en Astrophysique du Qu\'ebec (CRAQ), Qu\'ebec, QC, G1V 0A6, Canada\\
	$^{11}$Department of Physics and Astronomy, University of Hawaii at Hilo, Hilo, HI 96720, USA\\
	$^{12}$Aix Marseille University, CNRS, CNES, Laboratoire d’astrophysique de Marseille, 38 Rue Fr\'ed\'eric Joliot Curie, Marseille, 13013, France\\		
}

\date{Accepted XXX. Received YYY; in original form ZZZ}

\pubyear{\the\year{}}

\begin{document}
\label{firstpage}
\pagerange{\pageref{firstpage}--\pageref{lastpage}}
\maketitle

\begin{abstract}		
\noindent
As part of the SIGNALS survey, which comprises a sample of approximately 40 nearby galaxies observed with the Fourier transform spectrometer SITELLE, we present a study of metal mixing in the spiral galaxy NGC~6946.  
Taking advantage of the blue sensitivity of our setup, we measure the oxygen and nitrogen abundances of 638 \hii\ regions, and focus our analysis on the abundance fluctuations about the radial gradients. 
We detect an azimuthal variation of about 0.1 dex in these abundances across the NE spiral arm, with the leading edge being more metal-poor than the trailing edge. This result aligns with galaxy simulations, where radial gas flows along the spiral arms lead to dilution on the leading edge and enrichment on the trailing edge, due to the presence of radial metallicity gradients.
Our 2D analysis reveals that oxygen and nitrogen exhibit comparable spatial correlation scales, despite the different injection energies and distinct nucleosynthetic origins -- core-collapse supernovae in the case of oxygen and primarily AGB stars for nitrogen. 
The observed similarity suggests that 
stellar processes drive these two elements into the ISM over equivalent spatial scales.


\end{abstract}


\section{Introduction}

The existence of radial gradients in the present-day metallicity of spiral galaxies has been known and well-studied for decades from observations of their \hii\ regions (\eg\ \citealt{Searle:1971, Vila-Costas:1992, Zaritsky:1994, Moustakas:2010, Sanchez:2014, Zurita:2021}) and massive stars (\citealt{Urbaneja:2005, Urbaneja:2005a, Kudritzki:2008, Kudritzki:2012, Bresolin:2009a, Bresolin:2022}). On the other hand, gathering evidence for second-order metallicity variations within galaxy discs along the azimuthal coordinate has proven to be elusive using traditional long-slit spectroscopy (\citealt{Kennicutt:1996, Martin:1996, Bresolin:2011, Li:2013}) or other techniques (\citealt{Cedres:2002, Rosales-Ortega:2011}). The order-of-magnitude increase in the number of spectra collected within individual galaxies enabled by integral field spectroscopy 
has led to the unequivocal detection of azimuthal oxygen abundance gradients in spiral galaxies, in the form of differences on the order of $\sim$0.05-0.1 dex between arm and interarm regions, or across the widths of spiral arms  (\citealt{Sanchez:2015, Vogt:2017, Sanchez-Menguiano:2017, Ho:2017, Ho:2018}).
Classical chemical evolution models incorporating the effects of spiral arms on the distribution of metals, and oxygen in particular, have been introduced by \citet[see also \citealt{Spitoni:2023}]{Spitoni:2019} and \citet{Molla:2019}, who identified azimuthal fluctuations with amplitudes on the order of 0.1 dex, compatible in magnitude with those observed. Comparable results have been obtained from numerical simulations (\citealt{Solar:2020, Khoperskov:2023}).

Azimuthal variations in gas-phase metallicities are, however, not universally detected (\citealt{Zinchenko:2016, Williams:2022, Chen:2024}), and even when they are found, the association with the spiral arms or the galactic environment is not always manifest, with different authors drawing contrasting conclusions. For example, only half of the sample of spiral galaxies analyzed by \citet{Kreckel:2019} 
displays azimuthal variations of the \hii\ region metallicities, the link with the spiral arms sometimes missing, and thus providing no definitive indication for a systematic arm \vs\ interarm distinction. 
\citet{Sanchez-Menguiano:2020} instead conclude that the arms are more metal-rich than the interarm regions in about half of their sample, an effect that is reinforced in more massive and grand-design galaxies.  
Also \citet{Chen:2024} find that metallicity variations across arms, when detected, are larger in galaxies with more prominent spiral structures, while \citet{Grasha:2022} cannot establish a connection between azimuthal variations and spiral patterns.
A correlation between spiral arms and metal enrichment has also been detected in the Milky Way from samples of giant stars observed by {\it Gaia} (\citealt{Poggio:2022, Hawkins:2023, Barbillon:2025}). We can draw the conclusion that {\em observationally} it is still unclear whether the local physical conditions of the interstellar medium (ISM) 
or the existence of spiral arm structures represent the main drivers for the azimuthal distribution of the gas.

From the {\em theoretical} perspective the spiral arms appear to be crucial for sustaining the mixing processes that lead to the development of azimuthal chemical abundance patterns, either as conduits for large-scale radial motions of stars and gas  (\citealt{Grand:2016, Sanchez-Menguiano:2016, Orr:2023}) or as catalysts for the dilution of local metal self-enrichment due to the associated density waves (\citealt{Ho:2017, Molla:2019, Spitoni:2019}). 
In the former case the presence of a radial abundance gradient is required in order to develop azimuthal  variations, as elucidated by recent numerical simulations of Milky Way-type galaxies (\citealt{Orr:2023, Khoperskov:2023, Graf:2024}). These show how metal-poor gas flows along the arms from the galactic outer regions, while metal-rich gas moves in the opposite direction. Subsequent mixing gives rise to azimuthal metallicity variations. 
\citet{Orr:2023} find that the latter, at least in flocculent spiral galaxies, are not developed between the arm and interarm regions, but rather between arm structures. In the simulations by \citet{Grand:2016}, outward streaming stellar motions occur on the trailing edge of spiral arms, while inward motions take place along the leading edge, thus inducing a metallicity gradient across the arms (\ie\ azimuthally), characterized by a higher abundance on the trailing side (this occurs inside the corotation radius). \citet{Sanchez-Menguiano:2016} argue that this effect is consistent with the nebular oxygen abundance distribution they observe in the galaxy NGC~6754. Moreover, these authors detect radial gas motions along the arms, with a line-of-sight velocity amplitude of $\sim$30~\kms.
It is worth emphasizing that the local enrichment\,+\,mixing scenario proposed by \citet{Ho:2017}, put forward to explain the azimuthal abundance pattern observed in NGC~1365, produces higher metallicity at the location of the spiral arms, and lower metallicities on both the trailing and the leading sides. Such a pattern is different from what \citet{Sanchez-Menguiano:2016} detected in NGC~6754.

The study of the spatial distribution of metals in disc galaxies is closely connected with the description of metal transport mechanisms.
	Mixing processes, that lead to the homogenization of the metal field in the ISM, are largely regulated by interstellar turbulence (\citealt{Scalo:2004}), driven by the combined energy injection from stellar feedback, radial gas flows and cosmic accretion (\citealt{Sharda:2024}). \citet{Yang:2012} have shown that turbulent mixing can establish
kpc-scale homogenization in disc galaxies in less than an orbital timescale.	The effectiveness of turbulence
on non-axisymmetric metallicity distributions can explain the small azimuthal variations that are normally observed (\citealt{Petit:2015}).

Metal mixing in the ISM of spiral galaxies can be probed by evaluating the correlation length of the spatial distribution of the oxygen abundance fluctuations, using the two-point correlation function, as explained by \citet{Krumholz:2018}. These authors introduced a model based on stochastically forced diffusion, where the competing processes of metal injection and subsequent turbulent mixing can be treated analytically. This model predicts that in spiral galaxies the abundances of alpha elements such as oxygen (the most easily measured element from \hii\ region spectra), produced by core-collapse SNe, should be spatially correlated at the 50 (30) per cent level on scales of $\sim$0.5~kpc ($>$1~kpc).
The correlation scales measured by \citet{Kreckel:2020}, \citet{Williams:2022} and \citet[see also \citealt{Li:2023}]{Li:2021}  
are in substantial agreement with these predictions, despite the important simplifications made in the model, such as the neglect of spiral and bar structures.

In this paper we present a study of the azimuthal variations and the correlation function of oxygen and nitrogen abundances in NGC~6946, an isolated barred spiral galaxy at a distance of 7.72~Mpc (\citealt[1~arcsec = 37.4 pc]{Anand:2018}).
The addition of nitrogen in the analysis differentiates our work from previous investigations of the 2D chemical abundance distribution in spiral galaxies.

NGC~6946 has experienced a high rate of star formation over the recent past and a phenomenal rate of supernova explosions in historical times (\citealt{Botticella:2012, Eldridge:2019, Tran:2023}).
The galaxy displays a complex, multi-arm spiral system, due to the superposition of two- and three-arm structures (\citealt{Elmegreen:1992}).  
The spiral pattern is trailing, as evidenced by \citet{Sakhibov:2021}.
The `thick arm' to the NE of the galaxy nucleus warranted inclusion of NGC~6946 in Arp's Atlas of Peculiar Galaxies (\citealt{Arp:1966}). Integral Field Unit (IFU) observations of the star-forming complexes located in its eastern tip have been presented by \citet{Garcia-Benito:2010},  
and \citet{Tran:2023} estimated that this kpc-size region alone contributed to 5 per cent of the global star formation over the past 6~Myr.

This paper is organized as follows: we present the observations and the catalogue of \hii\ regions in NGC~6946 in Section~\ref{sec:observations}, and explain the methodology followed to measure the chemical abundances in Section~\ref{sec:chemical_abundances}. We look in detail at the azimuthal trends we derive across the NE arm for both the oxygen and nitrogen abundances in Section~\ref{sec:azimuthal}. In Section~\ref{sec:correlation} we use the two-point correlation function and an additional statistical tool, the semivariogram, to examine the scales of homogeneity in the 2D distribution of the two elements. We summarize our results in Section~\ref{summary} and present selected results from an alternative abundance diagnostic in the Appendix.

\section{Observations and H\,{\small \textbf{II}} region catalogue}\label{sec:observations}
We observed NGC~6946 with the imaging Fourier transform spectrometer SITELLE (\citealt{Drissen:2019}) at the 3.6m Canada-France-Hawaii Telescope on Maunakea, as part of the Star formation, Ionized Gas, and Nebular Abundances Legacy Survey (SIGNALS, \citealt{Rousseau-Nepton:2019}),
a project targeting $\sim$40 star-forming galaxies within approximately 10~Mpc. 
The whole optical disc of NGC~6946, out to its isophotal radius (\rtf\,=\,5.74~arcmin, \citealt{de-Vaucouleurs:1991}, corresponding to 12.9~kpc), fits into the $11\times11$~arcmin$^2$ field of view of the instrument, which produces datacubes with a pixel size of 0.32 arcsec pixel$^{-1}$ (12~pc pixel$^{-1}$). Considering the \fwhm\,$\simeq$\,1.0 arcsec seeing measured from our datacubes, the actual sampling is $\sim$37~pc, which compares favorably with the spatial resolution of popular galaxy surveys like CALIFA (\citealt{Sanchez:2012a}) or SAMI (\citealt{Bryant:2015}), 
and is similar to that being achieved for the nearest galaxies of the TYPHOON  (\citealt{Grasha:2022}) or PHANGS-MUSE (\citealt{Emsellem:2022}) surveys.
\begin{figure}
	\includegraphics[width=1\columnwidth]{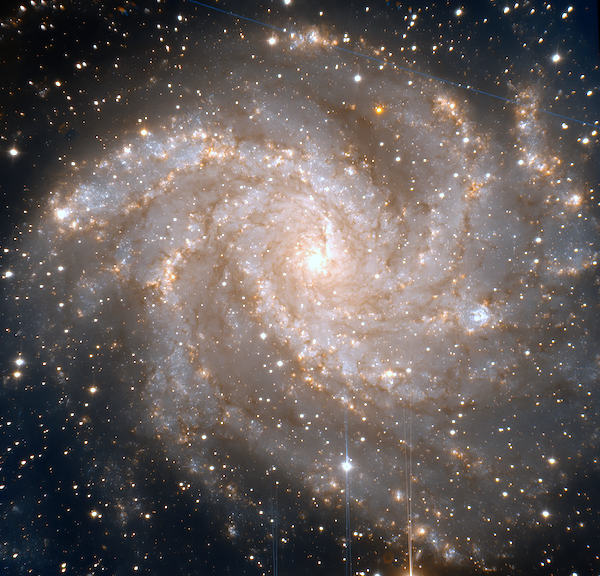}
	\caption{SITELLE view of NGC~6946 obtained from combining the SN2 and SN3 filter images.}
	\label{SITELLEimage}
\end{figure}  

The data were collected during nine separate nights in 2017 and 2018, with three different filter setups, SN1, SN2 and SN3, that isolate specific wavelengths. 
This choice gives access to the wavelength ranges 3640-3850\,\AA, 4840-5120\,\AA\ and 6480-6860\,\AA, respectively, where multiple nebular emission lines, useful for the chemical abundance analysis of the ionized gas, are located (Table~\ref{table:observing}). The instrument's throughput in the near-UV is critical to access the \oii\lin3727 line (actually the \llin3726-29 doublet), which is essential for our investigation.
Fig.~\ref{SITELLEimage} shows an image of NGC~6946 as seen by SITELLE, using a combination of the SN2 and SN3 filters, which highlights the multiple arm structure dotted with star forming regions.

\begin{table*}
		\caption{SITELLE observing parameters}\label{table:observing}
		\begin{tabular}{lccc}
			\hline
			& SN1	  & SN2	& SN3	\\
			\hline
			Dates					&		2018 Oct 5-11	&		2017 Sep 26-27	&		2017 Jul 01,03	\\
			Spectral range (\AA)	&		3640-3850		&		4840-5120		&		6480-6860		\\
			Exposure time per step (s)	&	59.0			&		50.2			&		36.1			\\
			Number of steps			&		240				&		203				&		323				\\
			Total integration time (h)	&		3.93		&		2.83			&		3.24			\\
			Spectral resolution		&		1000			&		1000			&		1900			\\
			Emission lines			&		\oii\lin3727	&		\hbeta, 		&	\halpha, \nii\llin6548,6583, \\[-0.5mm]
			&						&		\oiii\llin4959,5007	&	\sii\llin6717,6731		\\
			\hline
		\end{tabular}
	\end{table*}
	

In the off-axis interferometer lying at the core of SITELLE the optical path difference is modified by moving 
the scanning mirror in small steps, whose size determines the spectral resolution reported in Table~\ref{table:observing}. At each step a relatively small exposure is taken, and the interferogram datacube is the outcome of the full series of steps.

Spectral datacubes, calibrated in wavelength and flux, were produced by the fully automated data reduction pipeline {\sc orbs} (\citealt{Martin:2015, Martin:2021}). Refinements to the wavelength calibration of the higher-resolution SN3 datacubes, initially based on He-Ne laser observations, were carried out using sky emission lines. 

Further steps included the subtraction of the sky background and the correction for the stellar contribution.
To remove the sky background we used a median sky spectrum obtained from 12 regions of size 100$\times$100 spaxels located far away from the galaxy disc, rejecting spectra with intensity levels deviating from the average.
Due to the insufficient signal-to-noise ratio for detecting stellar absorption features, this study prioritizes other key analyses, with a detailed examination of stellar ages and metallicities left for future research.
Nevertheless, modelling the stellar contribution to the observed spectra remains crucial for mitigating the impact of absorption components on the Balmer emission lines, particularly in the case of \hbeta.
The methodology we follow, as previously presented by \cite{Rousseau-Nepton:2018}, utilizes a reference spectrum to subtract the contribution of the old stellar population. Such a spectrum is derived from a region devoid of emission lines and centred on the peak of the galaxy's continuum emission. After scaling this reference spectrum by the continuum level at each spaxel and adjusting for the local velocity, subtraction is performed, resulting in a pure emission data cube.

\subsection{H\,{\small \textbf{II}} region identification}

Observations made with SITELLE are ideal for describing the physical properties of ionized gas regions located in nearby galaxies. The large field of view of the instrument fully covers a galaxy as extended as NGC~6946 in a single pointing, and its high spatial resolution 
enables the observation of individual sites of star formation. 
In what follows we describe how such regions are defined and measured from the SITELLE data of NGC~6946.

The identification of gaseous ionized regions in galaxies typically makes use of continuum-subtracted narrowband (mostly \halpha) images and a variety of techniques, including percentage-of-peak photometry (\citealt{McCall:1996}), morphology-based object recognition as in the case of \mbox{HIIphot} (\citealt{Thilker:2000,Santoro:2022}), or fixed-threshold photometry (\citealt{Rozas:1999, Relano:2005}). 
More recently, codes optimized for the detection of \hii\ regions in integral field spectroscopic data have been introduced (\eg\ \citealt{Lugo-Aranda:2022}), and hierarchical clustering, a standard algorithm in astronomical data analysis (\citealt{Yu:2022}), has been employed (\citealt{Della-Bruna:2020, Rowland:2024}).


The second-generation \hii\ region detection code developed for SIGNALS is used here. Adapting the work of \citet{Rousseau-Nepton:2018} on NGC 628, with improved techniques described in Savard (2025, MSc, U. Laval, in prep.),
the code proceeds in three steps: (1) identification of the emission peaks; (2) determination of the zone of influence around each emission peak; (3) localization of the outer boundary of each region and the emission background, established by fitting a plane\,+\,2D Gaussian model.
Below we provide a succinct description of the main steps involved. 

\subsubsection{Peak identification}

Emission peaks are first identified in the Laplacian of the \halpha\ amplitude map, smoothed utilizing a Gaussian function with a standard deviation of 1.5 pixels (=\,0.48 arcsec) and considering a 3\,$\times$\,3 pixel detection box. These choices were dictated by the seeing during the observations, at the same time ensuring that peaks in crowded areas are properly differentiated. 
As described in the work of Savard (2025, in prep.), the detection threshold for the peaks is 
determined from the local background level (measured in a 6\,$\times$\,6 pixel box) and from its noise level. 


\subsubsection{Zone of influence}

The zone of influence (ZoI), describing the extent over which the observed flux can be traced to the ionizing photons from a single emission peak, is defined by assigning each pixel to the nearby peak that maximizes the value of $Amp(H\alpha)/r^2$, where the numerator is the \halpha\ amplitude of the peaks, and $r$ is the distance between these peaks and the selected pixel. Since ionized regions can vary widely in size, ranging from sub-parsec scales 
(as observed for ultra-compact \hii\ regions within the Milky Way, \citealt{Hoare:2005}) to several hundred parsecs in radius (giant \hii\ regions in the local Universe, \citealt{Kennicutt:1984}),
we adopted a maximum distance of 400 pc around each peak. 
In crowded regions this technique may create smaller, disconnected zones near the edges of the ZoI. These small islands, containing only a few pixels and negligible amounts of flux, are disregarded in the final ZoI before studying the emission spatial profile. 


\begin{figure*}
	\includegraphics[width=2\columnwidth]{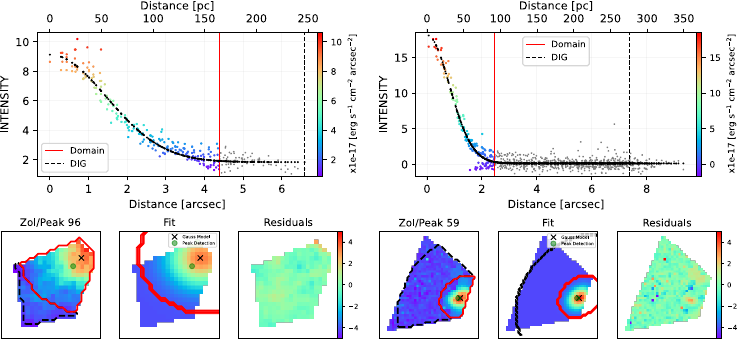}
	\caption[]{Zone of Influence (ZoI) for two selected peaks: the one on the left is in a crowded area, while the one on the right is more isolated. 
	{\em (Top panels)} The azimuthally averaged \halpha\ intensity profile is fitted with a plane\,+\,2D Gaussian with
	standard deviation $\sigma_{G}$. Here, the Gaussian profile is shown after the plane subtraction.
	The vertical red line is drawn at a radius equivalent to 3\,$\sigma_{G}$, marking the outer boundary of the Emission Region Domain (ERD).
	The Emission Background Domain (EBD) extends beyond this distance, out to a maximum of $9\,\sigma_{G}$. Its outer boundary is marked by the vertical dashed line.
	{\em (Bottom panels)} From left to right: \halpha\ map, plane\,+\,2D Gaussian model and residuals. The colour bars in the residuals plot map the \halpha\ intensity (same units as above).  The '$\times$' symbols indicate the emission peaks and the centers of the 2D Gaussian fits. The initial peak detections are shown by the green dots.  
	The red and black contours define the outer boundaries of the ERD and EBD, respectively. }
	\label{fit_regions_2d}
\end{figure*}

\subsubsection{Outer boundary and background emission}
\begin{figure*}
	\includegraphics[width=2\columnwidth]{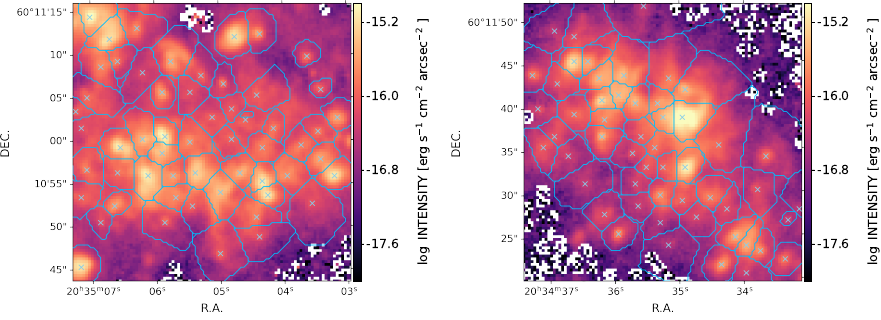}
	\caption[]{Segmentation maps of two randomly selected fields in NGC~6946, shown on top of the \halpha\ line emission. The '$\times$' symbols indicate the emission peaks, while the contours enclose the ERDs.}
	\label{_regions_segmentation}
\end{figure*}

A variety of techniques have been implemented in the literature to define \hii\ region boundaries, including emission line ratios, \halpha\ equivalent widths, spatial gradients of the \halpha\ flux, or a combination of the latter two (see \citealt{Santoro:2022}, and references therein). In this work, once the ZoI of a given peak has been identified, the Emission Region Domain (ERD) is established by fitting a 2D Gaussian (+ plane) profile of standard deviation $\sigma_{G}$ (average of the standard deviations along the two axes, $\sigma_x$ and $\sigma_y$)
to the pixels located within the ZoI. The procedure is illustrated for two different regions in Fig.~\ref{fit_regions_2d}.
By our definition, the ERD encloses all pixels located within $3\,\sigma_{G}$ of the peak (the radial distance marked by a vertical red line and labeled {\tt Domain} in Fig.~\ref{fit_regions_2d}). 
This can be taken as a proxy for the size of each region. In addition, an Emission Background Domain (EBD) is defined around each ERD as an annulus extending from 3\,$\sigma_{G}$ to a maximum of 9\,$\sigma_{G}$ if the radial extension of the ZoI allows it (the outer boundary is reduced in crowded regions, such as the one shown on the left of Fig.~\ref{fit_regions_2d}). 
The median intensity in this section defines the value of the background emission, which we assign to the 
presence of the diffuse ionized gas (DIG), and is subtracted from the flux integrated within the ZoI.
The vertical dashed line (labeled {\tt DIG} in Fig.~\ref{fit_regions_2d}) marks the outer boundary of the EBD.
In Fig.~\ref{_regions_segmentation} we display segmentation maps of two different fields in NGC~6946, generated by the technique outlined above.

\begin{figure*}
	\includegraphics[width=2\columnwidth]{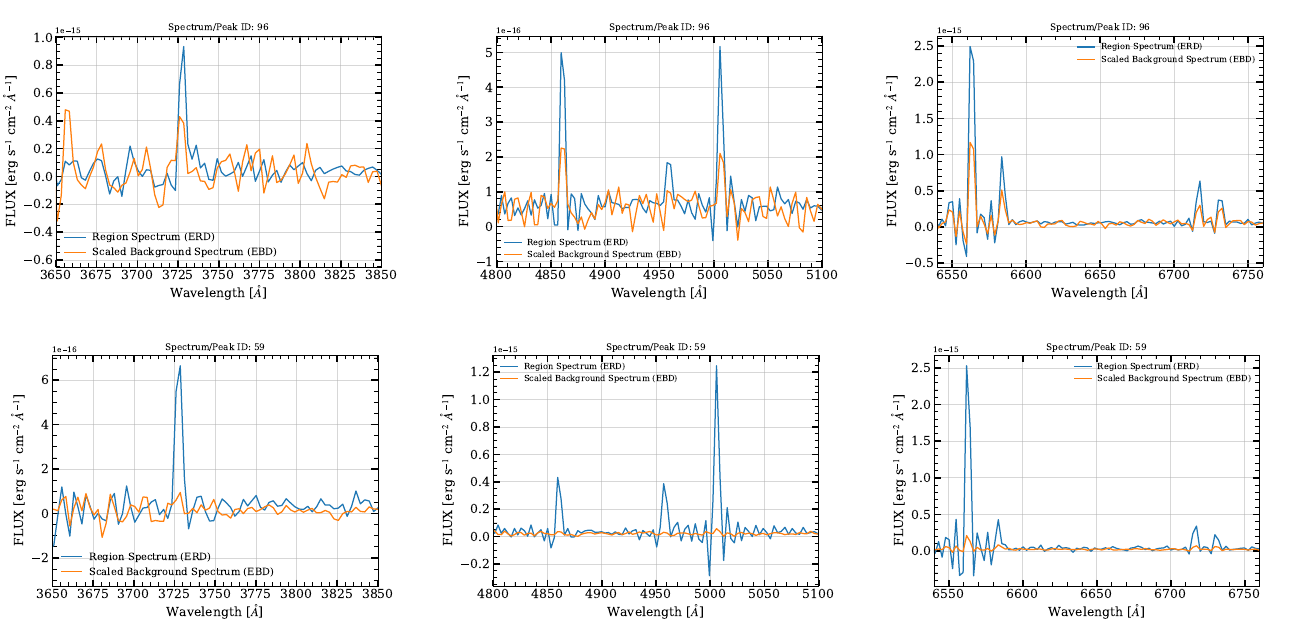}	
	\caption[]{Integrated spectra of the two regions in Fig.~\ref{fit_regions_2d}. The blue line represents the  spectrum extracted from the ERD, whereas the orange line represents the scaled background spectrum from the EBD. 
	}
	\label{fit_spectrum_2d}
\end{figure*}  

\begin{figure*}
	\includegraphics[width=1.7\columnwidth]{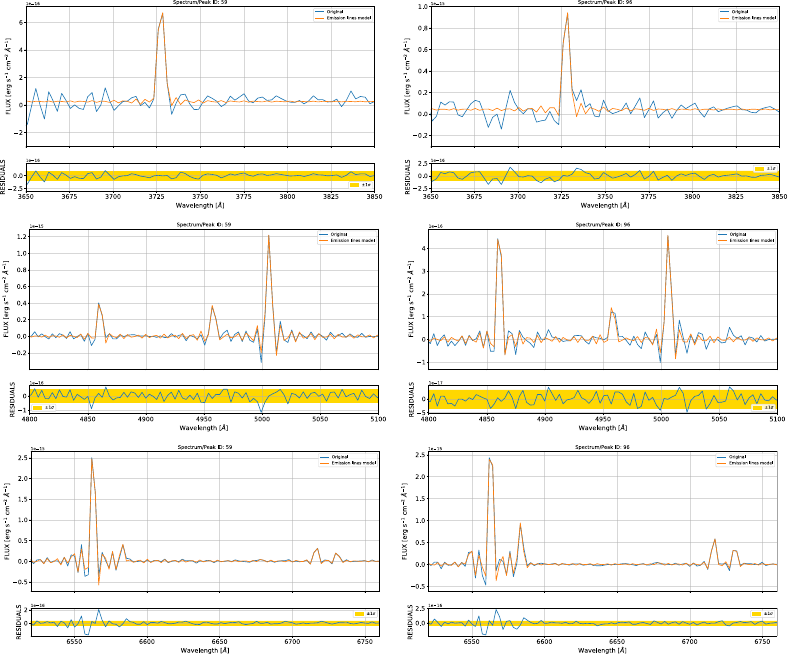}
	\caption{Spectra of the two regions in Fig.~\ref{fit_regions_2d}, after subtraction of the background. The blue line represents the final spectrum, while the orange line depicts the fit performed by \texttt{ORCS}.}
	\label{fit_spectrum_2d_2}
\end{figure*}  

\subsection{Classification and selection of integrated regions}

After identifying the regions, we measured the nebular emission features from the spatially integrated spectra (Fig.~\ref{fit_spectrum_2d})
with the \texttt{ORCS} post-processing software (\citealt{Martin:2015, Martin:2020}), which fits the lines using a sinc-Gaussian function, \ie\ the convolution of a sinc function (the instrumental line function of Fourier transform spectrometers) with a Gaussian (approximating the intrinsic line shape). 
Examples are presented in Fig.~\ref{fit_spectrum_2d_2}.

The line fluxes were corrected for the effect of dust extinction using the curve by \citet{Cardelli:1989}, assuming an electron temperature \te\,=\,10$^4$~K.
We then used classical BPT (\citealt{Baldwin:1981}) diagrams to remove regions that are inconsistent with photoionization by hot stars, adopting the diagnostic curves of \citet{Kauffmann:2003} in the 
log(\nii\lin6583/\halpha) \vs\ log(\oiii\lin5007/\hbeta) plane and
\citet{Kewley:2006} in the log(\sii\llin6717,6731/\halpha) \vs\ log(\oiii\lin5007/\hbeta) plane.

We required a signal-to-noise ratio SNR\,$>$\,4 for each of the emission lines used in the chemical abundance analysis (Section~\ref{sec:chemical_abundances}), leaving a total of 688 \hii\ regions out of 3509 detected peaks.
The bottleneck is represented by the \oii\lin3727 line, which suffers from 
high dust attenuation levels (1.48 magnitudes at 3500~\AA, \citealt{Schlafly:2011}),
due to the low Galactic latitude (11\fdg7 deg) of NGC~6946.


\section{Chemical abundances}\label{sec:chemical_abundances}

Since our aim is to study the 2D abundance distribution of both oxygen and nitrogen we require independent and uncorrelated abundance diagnostics for the two elements. This rules out methods anchored to photoionization models that incorporate prescriptions for the N/O\footnote{This notation denotes the abundance by number of the elements involved, \ie\ we write, for example, N/O instead of $n$(N)/$n$(O).} ratio as a function of O/H, motivated by observations. With reference to some of the studies mentioned in the Introduction, this is the case of the method followed by \citet{Ho:2017}, based on the Bayesian inference code {\sc izi} (\citealt{Blanc:2015}) and grids of photoionization models by \citet{Dopita:2013}, as well as the line diagnostic introduced 
by \citet{Dopita:2016} and used by \citet{Chen:2024} (an example of the possible shortcomings of this metallicity indicator is included in the Appendix). 
These and similar techniques can be used to infer the gas-phase metallicity (in practice the oxygen abundance) of extragalactic \hii\ regions, although it is important to be aware of the pitfalls and limitations imposed, for example, by  changes in the ionization parameter or the intrinsic N/O ratio.
Consequently, oxygen abundance indicators that combine oxygen and nitrogen line strengths are to be avoided, in order to circumvent spurious results introduced by variations in the N/O ratio (\citealt{Maiolino:2019, Schaefer:2020}).

Our choice of line indices falls naturally to  
\rtwothree\,=\,(\oii\lin3727 + \oiii\llin4959,5007)/\hbeta\ and N2O2 = log(\nii\lin6583/\oii\lin3727),
which are sensitive to O/H and N/O, respectively (\citealt{Pagel:1979, Pagel:1992, Perez-Montero:2009a}). The former is also influenced by the ionization parameter, which can be deduced from the O32 = \oiii\llin4959,5007/\oii\lin3727\ line ratio (\eg\ \citealt{McGaugh:1991}), but the latter is not, given the 
similarity of the \op\ and \np\ ionization potentials (\citealt{Peimbert:1969}).

The chemical abundance analysis is carried out with the Bayesian inference code NebulaBayes (\citealt{Thomas:2018}) and a large set of photoionization models calculated with {\sc cloudy} v23.01 (\citealt{Chatzikos:2023}). This combination was recently applied to the study of SITELLE data of NGC~628 by \citet{Garner:2025}, to which we refer for details. 
In short, spectral energy distributions at solar metallicity, calculated with Starburst99 (\citealt{Leitherer:2014}), are used as input to {\sc cloudy}, adopting the stellar atmospheres by \citet{Pauldrach:2001} and \citet{Hillier:1998} for massive stars and the Geneva stellar tracks with high mass loss (\citealt{Meynet:1994}). 
The photoionization models do not assume a prescription for N/O, but rather they are constructed by exploring a wide range in O/H, N/O and the ionization parameter. This choice allows us to address our investigation of the chemical abundance distribution in the disc of NGC~6946 in a self-consistent way.
The observed \rtwothree, N2O2 and O32 line ratios are employed as priors for NebulaBayes.

The \rtwothree\ index is degenerate: a fixed value of the parameter yields two possible O/H solutions, corresponding to an upper (metal-rich) and lower (metal-poor) branch.
\hii\ regions in the upper branch lie above certain values of the N2O2 and \nii\lin6583/\halpha\ line ratios (\citealt{McGaugh:1991, Kewley:2008}). Evaluating these criteria assigns the full \hii\ region sample to the upper, metal-rich branch. 
However, in the turnaround region of the log(\rtwothree) \vs\ log(O/H) diagram, located at log(\rtwothree) $\gtrsim$ 0.65, the quality of the fits produced by NebulaBayes rapidly worsens with increasing log(\rtwothree), as judged by the $\chi^2$ values. 
We decided then to impose the restriction $\chi^2 < 0.5$, which removed 7\% of the sample, leaving 638 regions with reliable chemical abundance measurements.
A catalogue of these \hii\ regions is presented in Table~\ref{table_region_fluxes}, where for each object we report celestial coordinates and fluxes, with the respective errors, as given by \texttt{ORCS}, for the following lines: \oii\lin3727, \hbeta, \oiii\lin5007, \halpha, \nii\lin6583, \sii\lin6717 and \sii\lin6731.

\begin{table*}
	
	\caption{Positions and emission line fluxes of the regions studied in NGC~6946, in units of $10^{-17}$ \uflux. We include here the first 10 rows, the complete table is available as part of the online supplementary material.}
	\begin{tabular}{lccccccccc}\hline\hline
		ID  & RA [Deg] & DEC [Deg]  & \oii\lin3727 & \hbeta& \oiii\lin5007 &\halpha & \nii\lin6583 &\sii\lin6717 &\sii\lin6731  \\
		& J2000 & J2000         &  & &  & &  & &  \\\hline	
		001 &   308.88342 &   60.12317 &     117.7 $\pm$    34.0 &     133.7 $\pm$     6.2 &      95.5 $\pm$     5.5 &     477.6 $\pm$     9.3 &      76.1 $\pm$     2.0 &      52.7 $\pm$     6.6 &       35.8 $\pm$     6.6 \\
		002 &   308.87651 &   60.18765 &     189.5 $\pm$    48.6 &      94.2 $\pm$     9.9 &      53.2 $\pm$     8.1 &     355.7 $\pm$    10.0 &      74.8 $\pm$     2.2 &      43.0 $\pm$     7.1 &       35.9 $\pm$     7.2 \\
		003 &   308.86973 &   60.15038 &     331.4 $\pm$    41.5 &     234.7 $\pm$    10.6 &     361.3 $\pm$    12.3 &     924.3 $\pm$    13.3 &     153.9 $\pm$     2.9 &      89.5 $\pm$     9.4 &       60.7 $\pm$     9.4 \\
		004 &   308.86560 &   60.16170 &     118.0 $\pm$    26.3 &     143.3 $\pm$     9.4 &     399.6 $\pm$    12.7 &     498.6 $\pm$     6.2 &      99.1 $\pm$     2.0 &      77.0 $\pm$     6.5 &       54.8 $\pm$     6.5 \\
		005 &   308.86354 &   60.19744 &     105.3 $\pm$    30.1 &      65.6 $\pm$     6.0 &      53.5 $\pm$     5.5 &     216.7 $\pm$     7.1 &      55.1 $\pm$     1.6 &      40.6 $\pm$     5.2 &       27.3 $\pm$     5.2 \\
		006 &   308.86352 &   60.16682 &     206.4 $\pm$    29.0 &     134.4 $\pm$     5.4 &      79.6 $\pm$     5.5 &     632.3 $\pm$    12.5 &     148.2 $\pm$     2.8 &     103.6 $\pm$     9.0 &       60.3 $\pm$     9.0 \\
		007 &   308.86147 &   60.17031 &     305.5 $\pm$    50.3 &     157.5 $\pm$     8.8 &     190.3 $\pm$     9.3 &     683.0 $\pm$    14.8 &     139.3 $\pm$     3.3 &     120.1 $\pm$    10.8 &       82.5 $\pm$    10.7 \\
		008 &   308.85979 &   60.16645 &     207.6 $\pm$    34.6 &     129.3 $\pm$     9.2 &     127.3 $\pm$     9.0 &     696.2 $\pm$    13.1 &     164.8 $\pm$     3.0 &     129.1 $\pm$     9.5 &       84.6 $\pm$     9.5 \\
		009 &   308.85975 &   60.16807 &     353.7 $\pm$    35.1 &     228.8 $\pm$     9.5 &     226.2 $\pm$     9.3 &    1030.4 $\pm$    19.7 &     238.5 $\pm$     4.5 &     194.6 $\pm$    14.4 &      126.0 $\pm$    14.3 \\
		010 &   308.85973 &   60.16879 &     236.8 $\pm$    29.5 &     149.8 $\pm$     6.4 &     147.2 $\pm$     6.3 &     643.4 $\pm$    12.7 &     141.5 $\pm$     2.9 &     112.3 $\pm$     9.2 &       72.3 $\pm$     9.2 \\
\hline
	\end{tabular}
	\label{table_region_fluxes}
\end{table*}%

\citet{Moustakas:2010} found that empirical calibrations can underestimate the “true" \te-based gas-phase metallicity by $\sim$0.2-0.3 dex, while theoretical calibrations, based on photoionization models, yield abundances that are too high by the same amount. Consequently, the absolute uncertainty in the nebular abundance scale is about 0.7 dex (\citealt{Kewley:2008}). \citet{Garner:2025} demonstrated that the approach we follow yields oxygen and nitrogen abundances that are compatible with these systematic uncertainties when comparing to those made with electron temperature measurements obtained as part of the CHAOS project (\citealt{Berg:2015,Berg:2020, Croxall:2015, Croxall:2016}).

\subsection{Radial gradients}\label{sec:radialgradients}
\begin{figure}
	\includegraphics[width=\columnwidth]{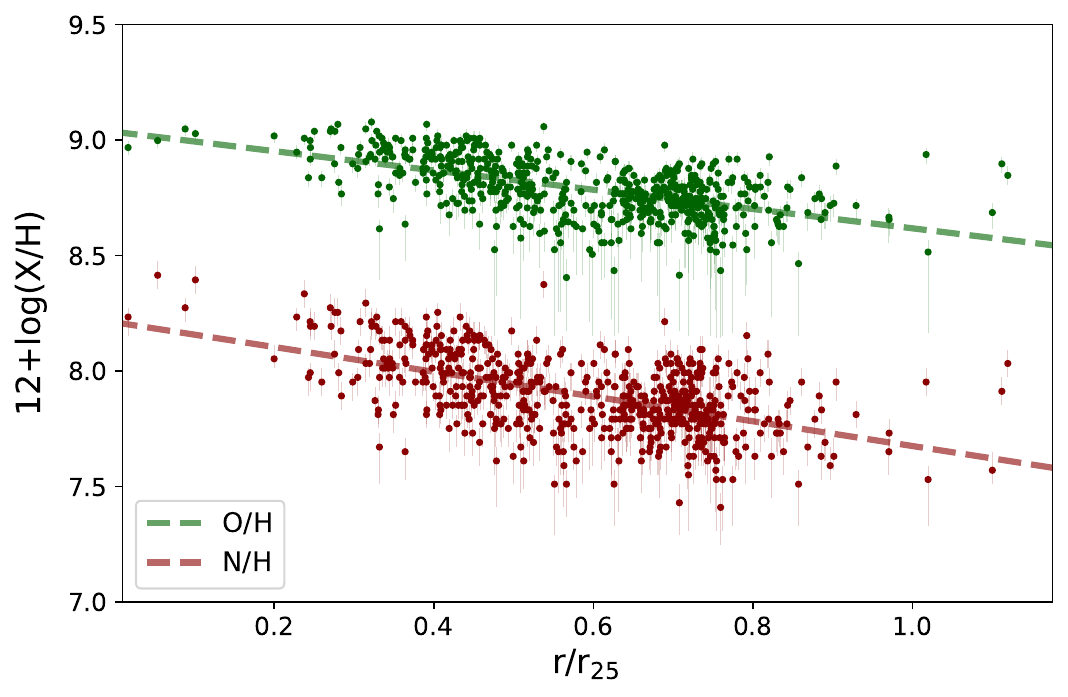}
	\caption{Oxygen (top, blue dots) and nitrogen (bottom, red dots) abundance gradients, with galactocentric distances normalized to the isophotal radius \rtf. The linear fits are displayed by the dashed lines.}
	\label{fig:gradient}
\end{figure}

Adopting the galaxy parameters in Table~\ref{table:parameters} we obtain 
the O/H and N/H radial gradients displayed in Fig.~\ref{fig:gradient}, where the galactocentric distances are normalized to the isophotal radius \rtf. 
The following expressions are linear fits to the data points:

\begin{equation}
{\rm 12 + log(O/H) = 9.04~ (0.01) - 0.42~(0.02) ~r/r_{25} }
\end{equation}
\begin{equation}
{\rm 12 + log(N/H) = 8.21~ (0.02) - 0.54~(0.03) ~r/r_{25}. }
\end{equation}

\noindent

\noindent
The exact functional form of these gradients is of secondary importance for the remainder of this work, which only takes abundance residuals relative to the radial trends into consideration. Because of the systematic differences affecting the nebular abundance diagnostics, mentioned above, the linear regression parameters and rms scatter about the fit can depend on the chosen abundance determination method, but our main conclusions are robust.
A potential flattening at large galactocentric distances cannot be ruled out (however, the number of data points is small), but since our analysis will be focused on the region $r<0.8$\,\rtf\ we do not attempt to account for it.
We note that the rms scatter measured about the linear regressions is 0.10 dex (for O/H) and 0.14 dex (for N/H), considerably larger than the values one measures adopting strong-line abundance diagnostics (see discussion in \citealt{Kreckel:2020}), but comparable to those obtained through the direct method (\eg\ \citealt{Croxall:2015, Berg:2015}). 
\citet{Lara-Lopez:2023} derived the oxygen abundance gradient in NGC~6946 from a large number of fiber spectra, adopting the S calibration by \citet{Pilyugin:2016}, and measured a scatter of 0.04 dex. Using the same technique we obtain an even smaller value, 0.03 dex.

\begin{table}
	
\centering
\caption{NGC 6946: galaxy parameters}\label{table:parameters}
\begin{tabular}{lcc}
\hline
RA (J2000)\phantom{aaaaaaaa}			&	20h 34m 52.33s			& NED \\
DEC (J2000)			&	60d 09m 14.1s			& " \\
i (degrees)			&	32.6				& \citet{de-Blok:2008} \\
PA (degrees)		&	242.7				& " \\
D (Mpc)				&	7.72				& \citet{Anand:2018}\\
\rtf\ (arcsec)		&	344.45				& \citet{de-Vaucouleurs:1991}\\
\hline
\end{tabular}

\end{table}
	

\section{Azimuthal variations of the gas-phase abundances}\label{sec:azimuthal}


\begin{figure*}
	\includegraphics[width=1.8\columnwidth]{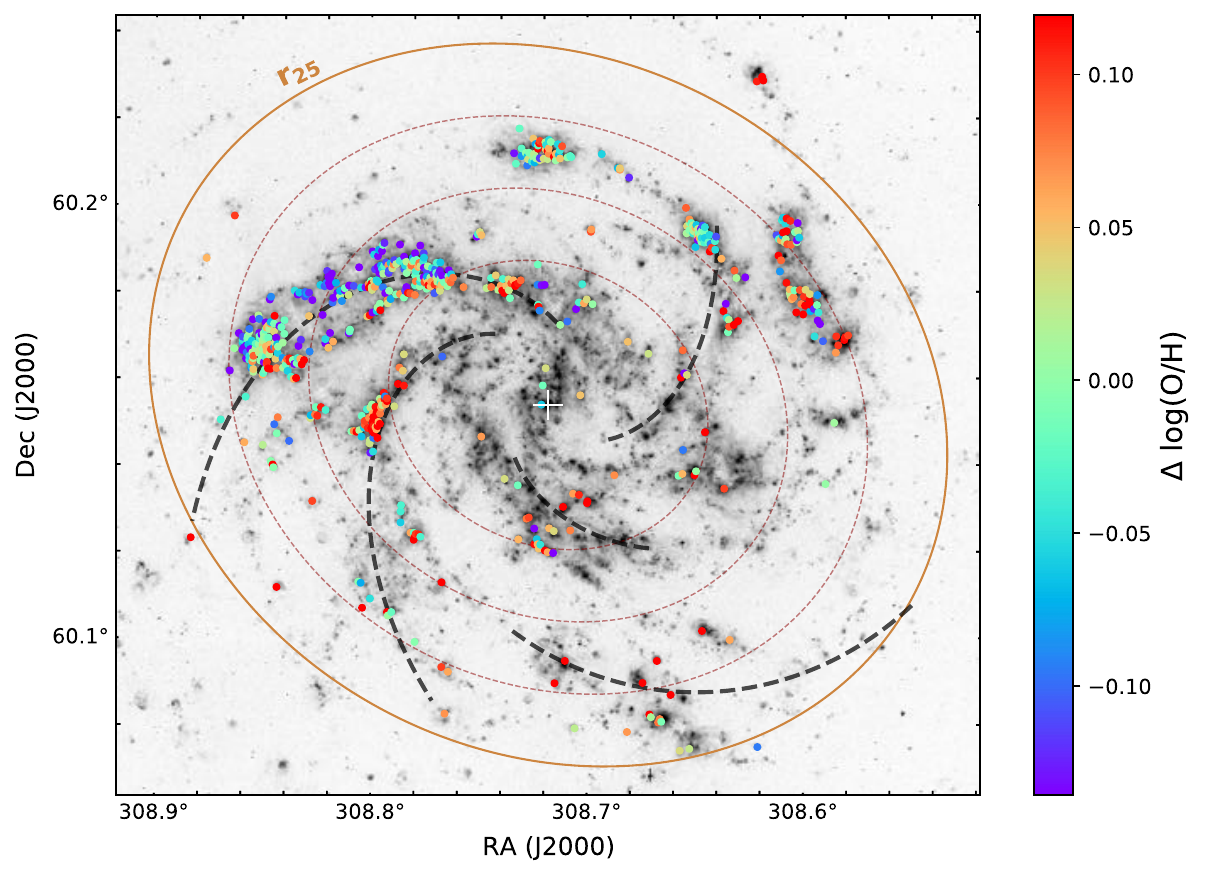}
	\includegraphics[width=1.8\columnwidth]{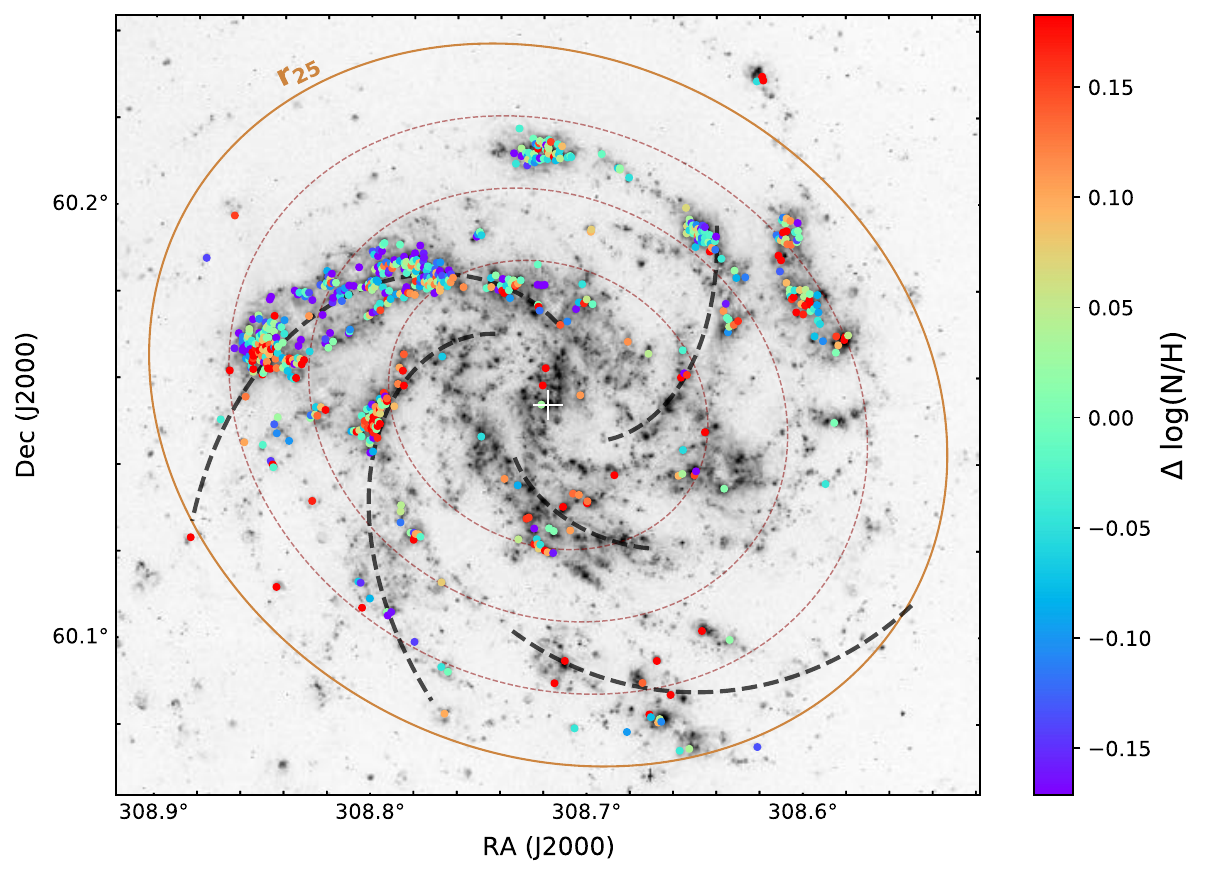}
	\caption{O/H (top) and N/H (bottom) residuals maps, shown on top of a narrow-band, continuum-subtracted \halpha\ image of NGC~6946 (William Herschel Telescope). For reference we display the five spiral arms identified in the optical $R$ band by \citet[dashed curves]{Frick:2000}, in addition to curves corresponding to the isophotal radius \rtf\ (outer  ellipse) and to 0.4, 0.6, 0.8 $\times$ \rtf\ (dotted ellipses).}
	\label{fig:fluctuations}	
\end{figure*}


Maps of the O/H and N/H abundance fluctuations about the galactocentric radial gradients are presented in Fig.~\ref{fig:fluctuations}: each dot corresponds to one \hii\ region, color-coded by its residual abundance. The two maps 
look qualitatively fairly similar, since the residuals $\Delta\log$(O/H) and 
$\Delta\log$(N/H) are well correlated.

It is evident that the vast majority of the \hii\ regions in our sample is located in the spiral arms, including the branches and fragments to the NW of the centre that were not traced in the optical $R$ band by \citet[dashed curves in Fig.~\ref{fig:fluctuations}]{Frick:2000}.
For this reason we do not attempt to carry out a comparison between the arm and interarm gas abundances (as done, for example, by \citealt{Kreckel:2019} and \citealt{Sanchez-Menguiano:2020}).
In addition, the NE arm (labeled R1 by \citealt{Frick:2000}) is the only structure that comprises a large number of high S/N regions over a significant spatial extent (about 6~kpc). Indeed, approximately half of the \hii\ regions in our final sample are associated with this structure. 
The NE arm displays a sizeable width on the sky (up to $\sim$60 arcsec, equivalent to $\sim$2.2~kpc), which facilitates the exploration of azimuthal effects across it. We will therefore limit the analysis in this section to this single feature. 
\citet{Cedres:2012} reported O/H azimuthal variations in NGC~6946, in the form of a high-metallicity structure to the S and W of the galaxy centre. We are unable to confirm their finding.

Before proceeding, we note that the thick portion of the NE arm is fully located inside the corotation radius ($\sim$0.8\,-\,0.9~\rtf: \citealt{Schinnerer:2006, Fathi:2007}). 
\citet{Sakhibov:2021} provide evidence that this structure rotates around the centre of the galaxy more slowly than stars and gas, as expected inside the corotation radius: in their clockwise motion around the galaxy centre, stars and gas enter the spiral arm at the trailing (southern) side.

\subsection{The NE arm}\label{NEarmanalysis}
A remarkable feature in Fig.~\ref{fig:fluctuations} is an apparent variation of the mean abundance residuals across the NE arm: on the trailing (southern) side they are on average larger than on the leading (northern) side. 
The amplitude of the effect, which at first glance is more discernible in the O/H map, is on the order of 0.1~dex.

\begin{figure}
	\includegraphics[width=1.\columnwidth]{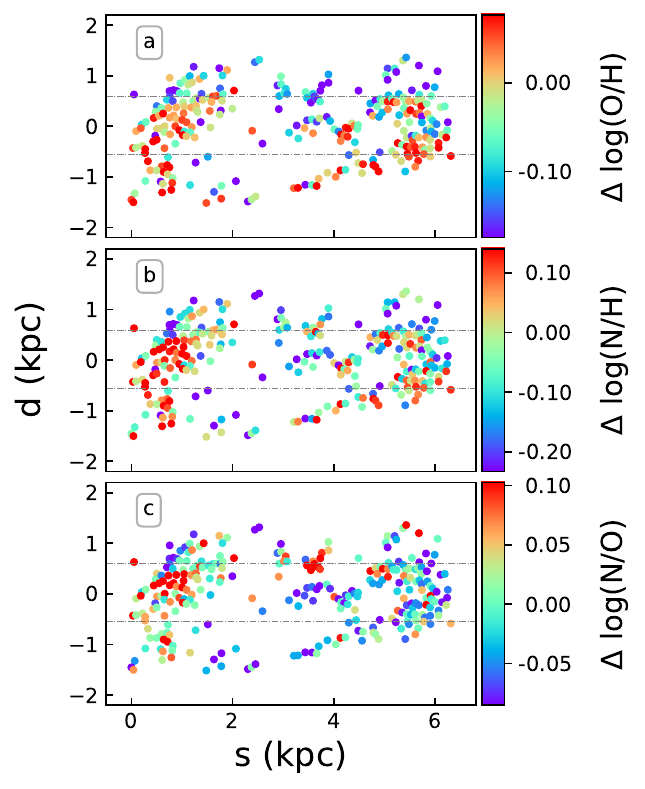}
	\caption{Residuals maps of (a) log(O/H), (b) log(N/H) and (c) log(N/O) for the NE arm \hii\ regions. The $s$ coordinate is the distance in kpc from the eastern tip of the arm, measured along the spiral arm. The $d$ coordinate is the distance in kpc above and below the central line of the spiral arm. The dashed horizontal lines mark the upper and lower 20 percentile levels of $d$.}
	\label{fig:NEarm}	
\end{figure}

In Fig.~\ref{fig:NEarm} we have isolated the \hii\ regions belonging to the thick part of the NE arm, extending approximately from 0.4~\rtf\ to 0.8~\rtf\ in distance from the centre, and unfolded the spiral structure on the plane of the galaxy. This figure displays the separation $d$ (in kpc) of the \hii\ regions from the arm's central line (\citealt{Frick:2000}) across the length of the arm, starting from the easternmost tip ($s$, in kpc).
We color-code the two top diagrams (panels a and b) by the values of the log(O/H) and log(N/H) residuals, obtained after subtraction of the galaxy radial abundance gradient. Panel c displays a map of the log(N/O) residuals.


We divide the sample into trailing ($d<0$) and leading ($d>0$) portions. 
A two-sided Kolmogorov-Smirnov (KS) test excludes the null hypothesis that the two sets are drawn from the same distribution, with a $p$-value $\ll$\,0.001 for O/H and equal to 0.003 for N/H. The smaller significance in the latter case can be attributed to the positive residuals on the leading side at the position of the star complexes 
at $(s,d)\simeq(0.8, 0.3)$~kpc. The N/O residuals map (Fig.~\ref{fig:NEarm}c) reveals a $\sim$0.1 dex enhancement at this very spot (a second region with a similar property appears at $s\simeq3.6$~kpc). This can potentially be explained by nitrogen enrichment of the ISM due to the pollution from winds of Wolf-Rayet stars (which have in fact been detected at this location by \citealt{Garcia-Benito:2010}). Although uncommonly, N/O enhancements attributed to these evolved massive stars have been reported in the literature (\eg\ \citealt{Kobulnicky:1997, Brinchmann:2008, Lopez-Sanchez:2010, Berg:2011, Karthick:2014}).

In order to quantify the abundance gradient azimuthally across the width of the arm we calculate the mean abundance difference between the \hii\ regions belonging to the top and bottom 20 percentiles of the $d$ values (indicated in Fig.~\ref{fig:NEarm}), selected to better represent nebulae located in the leading and trailing edges of the arm, respectively. This is of course an arbitrary choice, but it provides a rough measure of the effect we are investigating.
We find a decrease, moving from the trailing to the leading edge, 
of $\delta\log$(O/H)\,=\,0.11 (0.02) dex and $\delta\log$(N/H)\,=\,0.10 (0.03) dex: the azimuthal change in abundance is the same between the two elements.
The abundance dichotomy observed between the leading and trailing sides of the NE arm is consistent with the model predictions by \citet{Grand:2016} and the observation of a similar effect in NGC~6754 by \citet{Sanchez-Menguiano:2016}.
Recent galaxy simulations (\citealt{Orr:2023, Khoperskov:2023, Graf:2024}) indicate that inward and outward gas and stellar motions along the spiral arms generate an azimuthal abundance dispersion, and that the existence of radial abundance gradients is necessary for the development of azimuthal abundance variations: steeper radial gradients lead to more pronounced azimuthal variations. In our case the nitrogen radial gradient is only slightly steeper than that of oxygen (Sec.~\ref{sec:radialgradients}), and this is possibly the reason why 
the variation across the NE arm is similar between the two elements.

\citet{Chen:2024} detected O/H gradients across the spiral arms of NGC~1365 ($\sim$0.07~dex) and NGC~1566 ($\sim$0.04~dex). Their interpretation follows \citet{Ho:2017}, although 
the latter authors explain that their local enrichment model produces enhanced metallicities at the location of the spiral arms, in relation to both the trailing and the leading sides (see discussion in Section~5.4 of their paper). 
In any case, spiral arms do not typically display detectable evidence for chemical abundance azimuthal variations, and as summarized in the Introduction, observations suggest a variety of  scenarios.
Even if the interpretation offered by models and simulations is viable in the case of the NE arm of NGC~6946, as well as the arms of a few other galaxies, additional factors must play a role in determining whether azimuthal structures in the distribution of metals are developed. 




\section{Homogeneity scales of oxygen and nitrogen}\label{sec:correlation}

Whether oxygen and nitrogen are mixed into the ISM on similar spatial scales remains a poorly explored and outstanding question, although this issue is relevant for a deeper understanding of the processes affecting the chemical evolution of galaxies (\citealt{Emerick:2018}). The production sites differ between the two elements: oxygen originates from massive ($>$\,8~\msun) stars exploding as core-collapse supernovae, while nitrogen production is dominated by intermediate-mass (4-8~\msun) stars during the AGB phase (\citealt{Henry:2000, Kobayashi:2011, Vincenzo:2016, Vangioni:2018}). These stars are responsible for about 75\% of the nitrogen synthesized at solar metallicity (\citealt{Kobayashi:2020}).
Because of the different stellar lifetimes, the release of oxygen into the ISM occurs over a smaller timescale ($<$\,30~Myr) compared to nitrogen ($\sim$250~Myr for a single stellar population, \citealt{Johnson:2023}).

A first-order comparison of the mixing properties of O and N could in principle be based on the scatter observed about the radial gradients (\eg\ \citealt{Kennicutt:1996, Bresolin:2011, Pilyugin:2014, Croxall:2015}), but the azimuthal averaging removes essential clues that only a 2D analysis is able to provide. An approach 
that transcends the limitations imposed by the  simplest statistics, represented by the azimuthally averaged radial gradients, has been put forward by \citet[hereafter KT18]{Krumholz:2018}. These authors introduced an analytical model based on stochastically forced diffusion, that includes metal injection by stellar sources (supernovae, AGB stars and neutron star mergers) and turbulent mixing, approximated by linear diffusion. They emphasized the use of the two-point correlation of the 2D metal field as a measure of the spatial scales over which chemical elements are correlated, effectively the scales over which the distribution of the metals is
homogeneous. 
KT18 drew the conclusion that elements produced by core-collapse supernovae (alpha-elements, including oxygen) should be considerably more spatially correlated than those produced by AGB stars (s-type elements and, although not explicitly mentioned by the authors, nitrogen).
The correlation scales should vary between $\sim$1~kpc for the former, to $<$100~pc for the latter due to a smaller injection width, the spatial scale over which the metals are driven into the ISM. The reason lies in the different energetics involved in the release of the nucleosynthetic products: explosively in the case of SNe, and via low-velocity winds for AGB stars.

A first application of the two-point correlation function to gas-phase O/H abundance residual maps
was carried out by \cite{Kreckel:2020} in a sample of eight PHANGS-MUSE galaxies (\citealt{Emsellem:2022}), essentially confirming the $\sim$1~kpc scale predicted by the KT18 model for the mixing of oxygen.
Subsequent work by \citet{Li:2021, Li:2023} and \citet{Williams:2022} consolidated these results using larger galaxy samples, and looking for trends with galactic properties, such as stellar mass and star formation rate. \citet{Metha:2021} followed a different technique, based on the use of the semivariogram, to investigate the homogeneity of the observed metallicity fluctuations in PHANGS-MUSE galaxies. They found that the metallicities 
between neighbouring \hii\ regions separated by less than $\sim$1~kpc are well correlated, in substantial agreement with the other works cited above.

The quantification of the mixing scale is carried out in slightly different ways between authors: \citet{Kreckel:2020} and \citet{Williams:2022} use the length that corresponds to a fixed value of the two-point correlation function, chosen to be either 50 per cent or 30 per cent, while \citet{Li:2021, Li:2023} calculate the correlation length from a parametric fit to the KT18 correlation function model. The latter approach can provide a better understanding of the underlying physical processes, provided the model is correct, by disentangling the effects of the injection width and the correlation length on the shape of the correlation curve (\citealt{Li:2021}).

\subsection{Two-point correlation function}\label{Sec:2pcorr}

In this section we compare for the first time the two-point correlation of the gas-phase abundances of both oxygen and nitrogen in the same galaxy, and follow, for simplicity, the approach by \citet{Kreckel:2020} to quantify the mixing scale. 
As in \citet{Kreckel:2020}, our results refer to the distribution of metals in discrete \hii\ regions, rather than single spaxels as in \citet{Li:2021, Li:2023}. \citet{Williams:2022} followed both approaches, and found moderately consistent results. We share with the investigations based on PHANGS-MUSE data also the excellent physical spatial resolution: 30~pc in our case (40-70~pc in \citealt{Kreckel:2020}), compared to the hundreds of parsecs average resolution of the data analysed by \citet{Li:2021, Li:2023}.

We use the two-point correlation of the observed chemical abundances as a function of spatial scale $r$ (see KT18, \citealt{Kreckel:2020}):

\begin{equation}\label{xi1}
\xi(r) = \left\langle    \frac{ \overline{S_X(\boldsymbol{r_i})\, S_X(\boldsymbol{r_j})} - \overline{S_X}^2   }{\overline{  \Bigl(S_X - \overline{S_X}\Bigl)^2 } }     \right \rangle	
\end{equation}

\noindent
where $\boldsymbol{r_i}$ and $\boldsymbol{r_j}$ are the vector positions of the \hii\ regions where the metal field $S_X$ is measured,  the angle brackets indicate the average over all possible choices of $\boldsymbol{r_i}$, and the horizontal bars represent the average over all \hii\ regions. In our case $S_X$ refers to the logarithmic abundance fluctuations of either O/H or N/H, with a mean value of zero.
According to KT18 (see also \citealt{Li:2021}), Eq.~\ref{xi1} can then be simplified to


\begin{equation}\label{xi2}
	\xi(r) = \frac{\langle \overline{S_X(\boldsymbol{r_i})\, S_X(\boldsymbol{r_j}) }\rangle }{\sigma^2}	
\end{equation}

\noindent
where $\sigma^2$
is the variance of $S_X$. For the evaluation of the two-point correlation from Eq.~\ref{xi2} we calculate the product $S_{X,i}\,S_{X,j}$ for each pair of \hii\ regions ($i,j$) separated by the distance $r_{ij}$, and divide the full sample into 0.1~kpc-wide bins in $r_{ij}$ within the range 0-4~kpc.\footnote{We adapted the code by \citet{Williams:2022}, available at \url{https://github.com/thomaswilliamsastro/metallicity_gpr/tree/master}.} By construction, the correlation decreases from a maximum value $\xi(0)=1$ (each \hii\ region correlating perfectly with itself) to zero at very large separations. 
We execute 100 realizations by Gaussian sampling the log(O/H) and log(N/H) residuals uncertainties in order to estimate the 1\,$\sigma$ widths of the correlation curves. In addition, we run 100 realizations where we randomize the abundance residuals distribution in order to derive the curve corresponding to a lack of correlation between neighbouring \hii\ regions.

Two distinct approaches have been followed in the recent literature concerning the  separation between pairs in Eq.~\ref{xi1} and \ref{xi2}: \citet{Kreckel:2020} and \citet{Williams:2022} adopt \mbox{$\lvert \boldsymbol{r_i} - \boldsymbol{r_j} \rvert \leq r$}, while KT18 and \citet{Li:2021, Li:2023} utilize \mbox{$\lvert \boldsymbol{r_i} - \boldsymbol{r_j} \rvert = r$}. The adoption of the equality sign in the latter case is in line with other uses of the two-point correlation function in astronomy (\eg\ \citealt{Peebles:1993}). We utilize both definitions below, and show that they provide compatible outcomes.
		
\begin{figure}
	\includegraphics[width=1.\columnwidth]{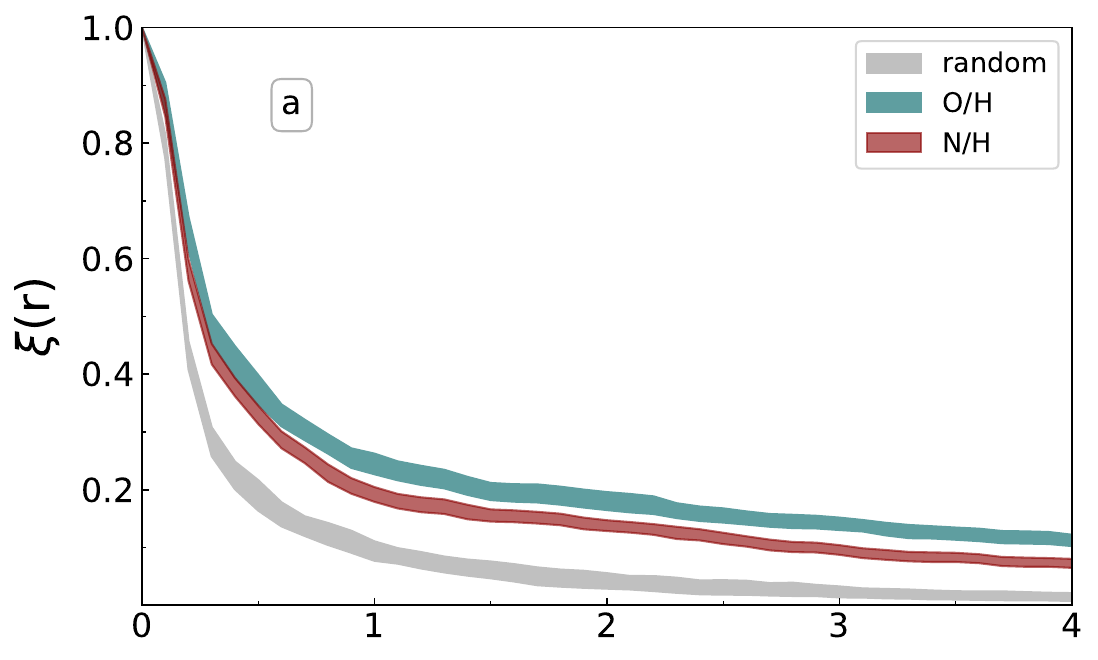}
	\includegraphics[width=1.\columnwidth]{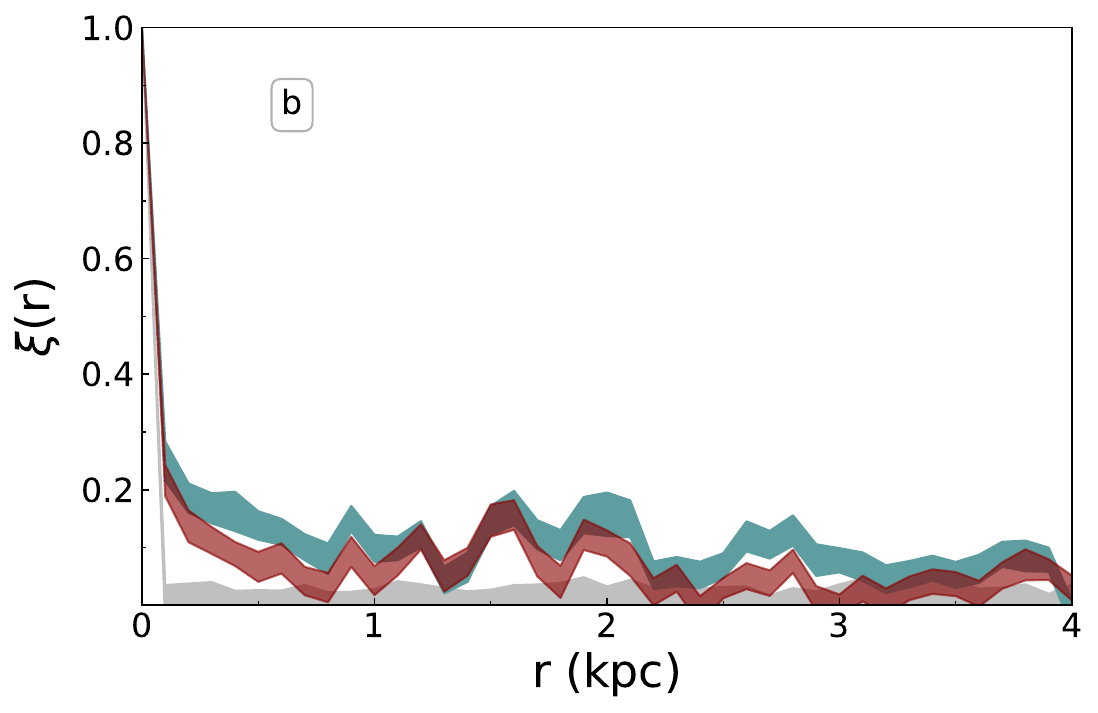}
	\caption{The two-point correlation functions of the oxygen (blue) and nitrogen (red) abundance fluctuations, using 0.1 kpc-wide bins. 
		The bands are drawn considering the 1\,$\sigma$ distribution of 100 realizations obtained by Gaussian sampling the uncertainties. The grey band is drawn from 100 realizations of a random shuffle of the abundances. We adopt two different definitions for the separations between pairs in Eq.~\ref{xi2}: (a) $\lvert \boldsymbol{r_i} - \boldsymbol{r_j} \rvert \leq r$, and  (b) $\lvert \boldsymbol{r_i} - \boldsymbol{r_j} \rvert = r$.
	}
	\label{fig:2Pcorrelation}	
\end{figure}

\smallskip
(a) Our results adopting $\lvert \boldsymbol{r_i} - \boldsymbol{r_j} \rvert \leq r$	
are displayed in Fig.~\ref{fig:2Pcorrelation}a, where it is evident that the curves (actually bands) for both oxygen and nitrogen deviate significantly from the curve obtained by randomizing the abundances, and that the N/H correlation curve lies consistently below the O/H curve. To estimate the significance of the offset, we randomly generated $10^4$ curve pairs (one curve for O/H and one for N/H) lying within the respective lower and upper bounds, each time calculating their offset along the $y$-axis, performing the KS test and deriving a $p$-value from the test. Since the curves touch each other at $r=0$, we avoided separations $r<0.2$~kpc. 
We find that the mean $p$-value is 0.026 with a standard deviation of 0.019. This implies that the offset seen in Fig.~\ref{fig:2Pcorrelation} can be considered as statistically significant, \ie\ the nitrogen correlation is slightly smaller than that of oxygen.
 

As a measure of the mixing scale of oxygen and nitrogen we interpolated the upper and lower bounds of each band to estimate the separation in kpc where the correlation falls to 50 per cent and 30 per cent:
\bigskip

\begin{tabular}{llc}
	& & $r$ (kpc)\\[2mm]
	50 per cent\phantom{aaa} 	&  O/H\phantom{aaa}	&	$0.29^{+0.03}_{-0.02}$\\[1mm]
	           					&  N/H				&	$0.26^{+0.01}_{-0.01}$\\[3mm]
	30 per cent 				&  O/H				&	$0.70^{+0.09}_{-0.09}$ \\[1mm]
               					&  N/H				&	$0.56^{+0.04}_{-0.04}$ \\	           
\end{tabular}

\bigskip\noindent
The correlation scales that we measure fall within the range derived (in the case of oxygen) for other nearby galaxies (see comparison in \citealt{Li:2021}).
The values for O/H and N/H differ only at the 1.5\,$\sigma$ level, which is less significant than that of the overall offset between the curves, because these scales are measured on the steep, descending portions of the curves themselves. The difference is only appreciated at large separations.

\smallskip
(b) If we use the standard definition of the two-point correlation, where 
$\lvert \boldsymbol{r_i} - \boldsymbol{r_j} \rvert = r$, we obtain the curves shown in 
Fig.~\ref{fig:2Pcorrelation}b, which look different from those illustrated above. In particular, they drop sharply at small separations and display significant fluctuations  in amplitude, due to the stricter criterion that isolates a much smaller number of H\,II regions per bin. The correlation function values are also considerably lower than in the previous case -- such low values are compatible with those calculated by \citet[see their Fig.~3]{Li:2021}. Nevertheless, the location of the N/H curve below the O/H curve is retained.

\smallskip
Although the trend we see in Fig.~\ref{fig:2Pcorrelation} goes in the direction expected by KT18, we do not find evidence for N/H to be characterized by a much smaller correlation scale ($<$\,100~pc at the 50 per cent level) compared to O/H, as indicated by the KT18 model with the adopted parameters (see their Fig.~4). Our result suggests that adjustments to the KT18 model parameters are called for, in particular with regard to the small injection width of AGB stars. We argue that this is justified by the significantly longer timescale over which AGB stars release their nucleosynthetic products into the ISM, in comparison to massive stars. In a single stellar population, such a timescale is on the order of 250~Myr, over which migration of the stars is likely to take place (\citealt{Johnson:2023}), thus leading to a large spatial extent over which nitrogen can be injected into the ISM.

\subsection{Semivariogram}
In order to consolidate our conclusions, we also explored the use of the semivariogram, a geostatistical tool that has been recently applied to the analysis of the 2D chemical abundance distributions in galaxies by \citet{Metha:2021, Metha:2022}. In our case the semivariogram $\gamma(r)$ measures how the variance of the chemical abundances changes as a function of separation between pairs of H\,II regions. Like the two-point correlation function, which is a strictly associated concept, it can inform us on the spatial scales of homeogeneity of the chemical abundance fields.
Using the quantities defined in Eq.~\ref{xi1}:
\begin{equation}
	\gamma(r) = \frac{1}{2}\, \mathrm{Var}\, (S_X(\boldsymbol{r_i}) - S_X(\boldsymbol{r_j}))
\end{equation}
where the variance is calculated over all pairs of H\,II regions ($i,j$) separated by a distance $r_{ij}$ such that $r-\delta/2 \leq r_{ij} \leq r + \delta/2$, with $\delta$  the chosen bin size (\citealt{Metha:2021}). 

\begin{figure}
	\includegraphics[width=1.\columnwidth]{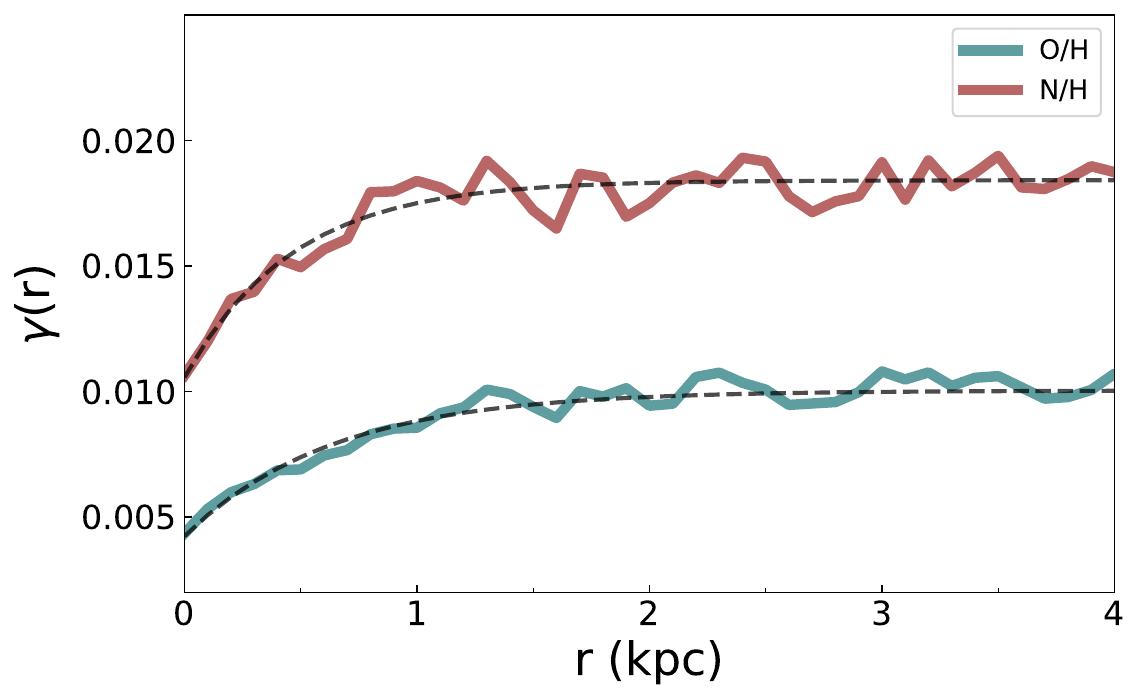}
	\caption{Semivariograms of the oxygen (blue) and nitrogen (red) abundance distributions. The dashed lines represent the fits to the model in Eq.~\ref{Eq:model}.
	}
	\label{fig:semivariogram}	
\end{figure}

The semivariograms for both log(O/H) and log(N/H), adopting \mbox{$\delta=0.1$}~kpc, are shown in Fig.~\ref{fig:semivariogram}. The vertical offset between the two is due to the larger (uncorrelated) measurement uncertainties of the N/H abundances, since the value $\gamma(r=0)$ increases with the measurement errors (\citealt{Metha:2021}). In our sample, the mean errors are 0.05~dex for log(O/H) and 0.07~dex for log(N/H). The observed semivariances at $r=0$ are considerably larger (by a factor of $\sim$\,3.5-4) than what is derived from the mean errors, suggesting the presence of correlated abundance fluctuations on smaller spatial scales.

The amplitude of the chemical abundance fluctuations can be estimated by the difference in $\gamma(r)$ measured between large separations, where the abundances are uncorrelated and the curve flattens out, and small separations. Such a difference is $\Delta\gamma\simeq$ 0.006-0.008 dex, implying fluctuations on the order of 0.11-0.12~dex.

In order to estimate the spatial scales over which the chemical abundances can be considered to be correlated, we followed \citet{Metha:2022} and fitted the semivariograms with a simple model:
\begin{equation}\label{Eq:model}
	\gamma(r) = A^2\,(1-exp(-r/R_S)) + \gamma_0	
\end{equation}	
where $\gamma_0$ is the offset measured at $r=0$, $A^2$ corresponds to the $\Delta\gamma$ value described above, and $R_S$ is a measure of the homogeneity scale of the chemical abundances. Our fits are shown as dashed lines in Fig.~\ref{fig:semivariogram}. We estimated the errors in the fit parameters by varying the range in $r$ over which the fit is performed (5-20~kpc in 0.1~kpc steps). We obtain $R_S=0.65\pm0.08$~kpc for log(O/H) and $R_S=0.47\pm0.03$~kpc for log(N/H). The marginal difference is consistent with our previous finding (Sect.~\ref{Sec:2pcorr}) that the correlation scale of N/H is slightly smaller than that of O/H. We calculate that the correlation functions for O/H and N/H shown in Fig.~\ref{fig:2Pcorrelation}a fall to $\sim$33 per cent at a separation $r=R_S$.

\medskip
In summary, using different statistical tools, we find evidence that chemical abundance fluctuations on the order of $\sim$0.1~dex exist in NGC~6946 for both oxygen and nitrogen. Although nitrogen is slightly less correlated than oxygen, the spatial correlation is comparable between the two elements, and is on the order of half a kpc.


The latter conclusion is supported by the results of recent high-resolution simulations of a Milky Way-type galaxy presented by \citet{Zhang:2024}, who tracked seven different chemical elements, including O and N, following  their injection into the ISM on a star-by-star basis. \citet{Zhang:2024} find that all the elements they considered are well correlated with each other (C representing a mild exception), and that N has virtually the same spatial statistics of the s-process elements Ba and Ce, generated by AGB stars, and deviating only marginally from that of O, S and Mg, released by core-collapse supernovae. N is found to have a slightly longer correlation length than O. The much larger injection width for N (and other AGB-related species) compared to the KT18 model is justified by the longer lifetimes of intermediate-mass stars, which implies longer timescales for the release of their nucleosynthetic products as well as a drift from their birth sites. This is in overall agreement with our findings and arguments, even though we measure a slightly smaller correlation for N compared to O, rather than the opposite effect as found by \citet{Zhang:2024}.

\section{Summary and conclusions}\label{summary}
The efficient mapping of the 2D distribution of metals across galaxy discs requires instruments capable of providing simultaneous spatial and spectroscopic information across a wide wavelength range. The availability of hundreds or thousands of spectra (of either spaxels or individual \hii\ regions) per galaxy 
enables the investigation of higher-order statistics of the distribution of metals compared to the mere derivation of abundance gradients. 

Due to instrument construction and sensitivity constraints, the extragalactic surveys carried out in the past decade have focused on deriving the distribution of the oxygen abundance alone, using a variety of line diagnostics and photoionization modeling. Broadening the variety of chemical species probed by the observations opens up new opportunities for the exploration of processes regulating the chemical evolution of galaxies. In this work, we have presented an analysis of the 2D distribution of oxygen and nitrogen in the disc of NGC~6946, made possible by the unique blue sensitivity and wide field of view of the SITELLE Fourier transform spectrometer. For the chemical abundance analysis we relied on the NebulaBayes code (\citealt{Thomas:2018}) and a custom grid of photoionization models (\citealt{Garner:2025})
that enable the exploration of oxgen and nitrogen abundance variations over a broad range, rather than adopt a prescribed N/O behaviour with O/H. This approach allows us to circumvent potential pitfalls intrinsic to strong-line diagnostics. Our analysis focuses on the abundance fluctuations obtained by subtracting the galactocentric gradients from the abundances.
We obtain two main results: \medskip

\noindent
(a) There is a $\sim$0.1 dex gradient in the residuals of both the O/H and N/H abundance ratios measured azimuthally across the NE arm of the galaxy, the only structure of the galaxy where we have sufficient statistics for the study of azimuthal variations, with the leading side of the arm being more metal poor than the trailing side. Such a trend matches the expectations from galaxy simulations (\citealt{Grand:2016, Sanchez-Menguiano:2016}), 
in which azimuthal variations are produced by the radial movement of gas and stars along the spiral arms, inwards (outwards) along their leading (trailing) edges. In the simulations, inward-bound gas flows dilute and lower the abundances of the gas, in virtue of the presence of radial gradients in the distribution of metals (\eg\ \citealt{Orr:2023}).
\medskip

\noindent
(b) Despite the differences in energetics and nucleosynthetic origin, with oxygen released explosively by massive stars  and nitrogen via slow winds by intermediate-mass stars, these two chemical elements appear to be distributed on similar spatial scales in the ISM of NGC~6946, with nitrogen marginally less correlated than oxygen. This goes against the much smaller correlation expected for nitrogen from the model by \citet{Krumholz:2018}, which stimulated our correlation analysis, but is in rough agreement with the findings from the galaxy simulations by \citet{Zhang:2024}.

\medskip


Of course, before one can confidently claim that our result concerning the correlation scales indicates drawbacks in the KT18 framework, and in particular the choice of the metal injection parameters, or, alternatively, issues with our observations and data analysis, will require the investigation of additional galaxies. The SIGNALS survey is well equipped, both in terms of instrumentation and galaxy sample size, to address the observational side of this issue.


\bigskip
\bigskip
\noindent

\section*{Acknowledgments}
Based on observations obtained with SITELLE, a joint project between Université Laval, ABB-Bomem, Université de Montréal and the CFHT with funding support from the Canada Foundation for Innovation (CFI), the National Sciences and Engineering Research Council of Canada (NSERC), Fond de Recherche du Québec - Nature et Technologies (FRQNT) and CFHT.
The authors wish to recognize and acknowledge the very significant cultural role and reverence that the summit of Maunakea has always had within the indigenous Hawaiian community. We are most fortunate to have the opportunity to conduct observations from this mountain.
This research has made use of the NASA/IPAC Extragalactic Database (NED), which is operated by the Jet Propulsion Laboratory, California Institute of Technology, under contract with the National Aeronautics and Space Administration. 
We thank Jefeng Li for his comments on our calculation of the two-point correlation function in an earlier draft.

DFA  acknowledges the support from the National Science Foundation under grant 2109124 for  SIGNALS: Unveiling Star-Forming Regions in Nearby Galaxies. LRN is grateful to the National Science Foundation NSF - 2109124, the Dunlap Institute, and the Natural Sciences and Engineering Research Council of Canada NSERC - RGPIN-2023-03487 for their support. The Dunlap Institute is funded through an endowment established by the David Dunlap family and the University of Toronto.
AZ acknowledges support from projects PID2023-150178NB-I00 and PID2020-114414GB-I00, financed by MCIN/AEI/10.13039/501100011033 and from the Junta de Andaluc\'ia (Spain) local government through the FQM-108 project.
CR and LD are thankful for a FRQNT-SIGNALS team grant and NSERC-Discovery grants.

This work made use of {\tt APLpy} (\citealt{Robitaille:2012, Robitaille:2019}), {\tt Matplotlib} (\citealt{Hunter:2007a}), {\tt NumPy} (\citealt{Harris:2020}), {\tt SciPy} (\citealt{Virtanen:2020}), NebulaBayes (\citealt{Thomas:2018}), {\sc cloudy} (\citealt{Chatzikos:2023}).

\section*{Data availability}
The data underlying this article will be shared on reasonable request to the first author.

\bibliographystyle{mnras}
\input{paper.bbl}

\appendix
\section{Results from an alternative metallicity diagnostic}

We include here the O/H residuals map obtained using the metallicity diagnostic by \citet{Dopita:2016}, which is based on the \nii\lin6583/\halpha\ and \nii\lin6583/\sii\llin6717,6731\ line ratios:

\begin{equation}\label{Eq:oh}
\rm 12+log(O/H) = 8.77 + y + 0.45(y+0.3)^5
\end{equation}
\noindent
where
\begin{equation}\label{Eq:y}
\rm y = \log([N\,II]/[S\,II] + 0.264\log([N\,II]/H\alpha).
\end{equation}
\smallskip

\begin{figure*}
	\includegraphics[width=2\columnwidth]{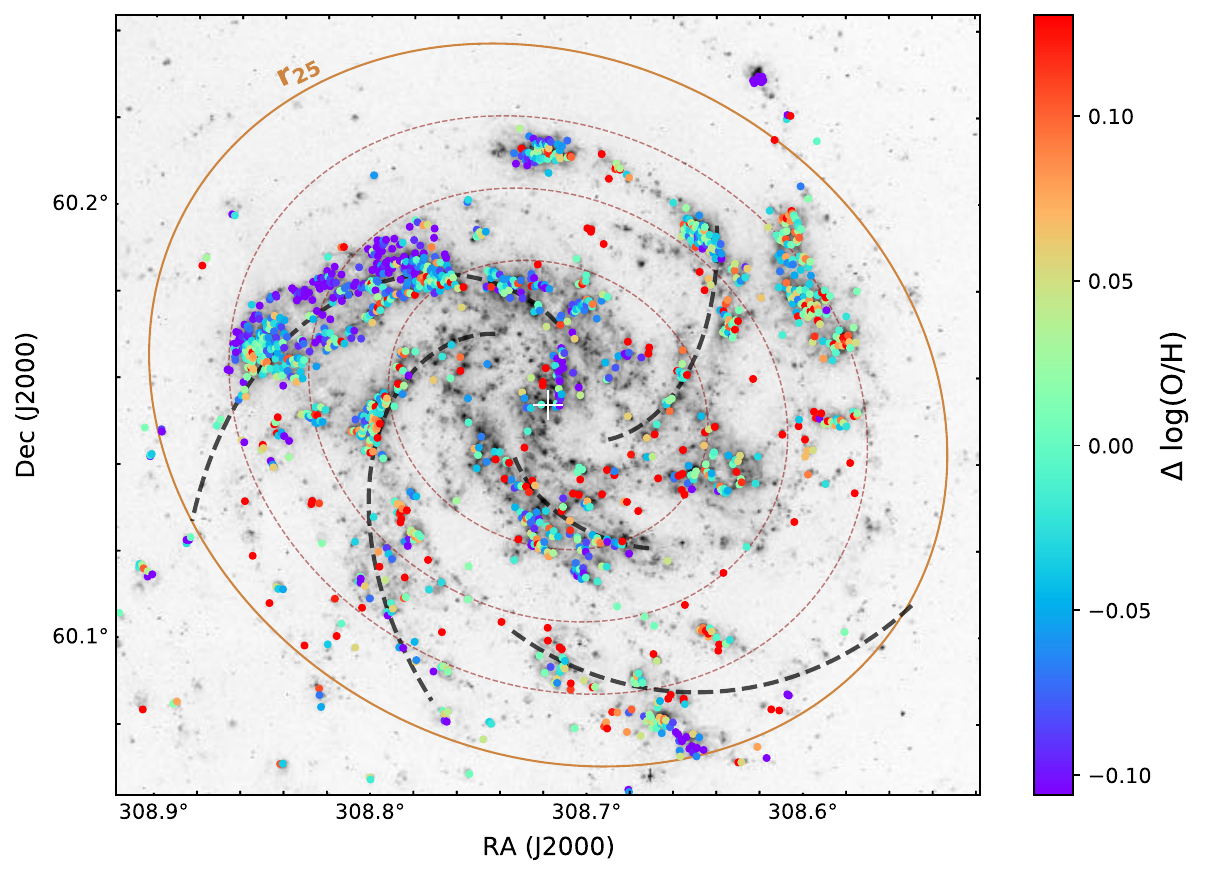}
	\caption{O/H residuals map, similar to Fig.~\ref{fig:fluctuations}, but using the metallicity diagnostic by \citet{Dopita:2016}.}
	\label{fig:D16map}	
\end{figure*}

\noindent
In this case our adopted signal-to-noise ratio contraint (SNR\,$>$\,4) is much less restrictive than in the main paper, since we are avoiding the \oii\lin3727 line, which is often not observed with a sufficient signal-to-noise ratio due to the high interstellar extinction, and we can rely on approximately 1400 metallicity data points distributed across the disc of NGC~6946. Of course, our abundance analysis remains limited to O/H.
The outcome is shown in Fig.~\ref{fig:D16map}.
It can be seen that the overall map is similar to the O/H map of Fig.~\ref{fig:fluctuations}, and in particular the abundance gradient across the NE arm is evident.

\begin{figure*}
	\includegraphics[width=1.2\columnwidth]{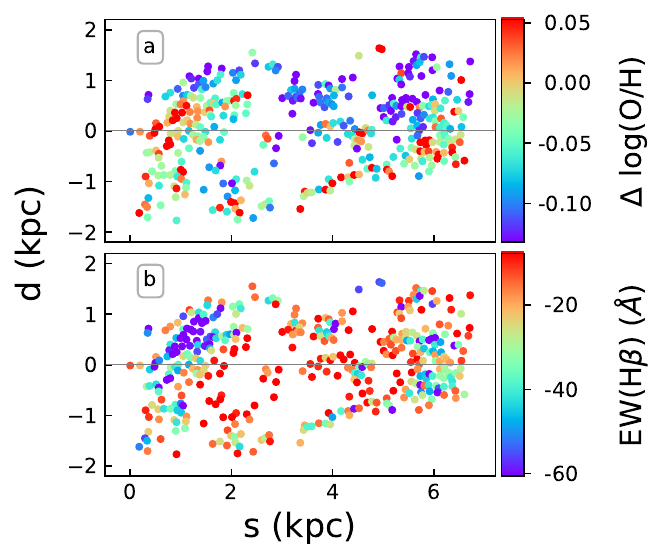}
	\caption{Maps of (a) the O/H residuals (adopting the \citealt{Dopita:2016} metallicity diagnostic) and (b) the \hbeta\ equivalent width for the NE arm \hii\ regions. The $s$ coordinate is the distance in kpc from the eastern tip of the arm, measured along the spiral arm. The $d$ coordinate is the distance in kpc from the central line of the spiral arm.}
	\label{fig:D16_NEarm}	
\end{figure*}

In Fig.~\ref{fig:D16_NEarm} we repeat the analysis of the NE arm we carried out in Section~\ref{NEarmanalysis}, but we substitute panels b and c of Fig.~\ref{fig:NEarm} (that refer to nitrogen) with a map of the \hbeta\ equivalent widths, EW(\hbeta). This is a proxy for the age of the \hii\ regions, and it indicates younger ages ($\simeq$\,4~Myr using the calibration of \citealt{Terlevich:2004} -- consistent with age inferred from the presence of Wolf-Rayet stars, detected by \citealt{Garcia-Benito:2010}) in the eastern star-forming complexes, compared to the rest of the arm ($\simeq$\,5-6~Myr).
Between the trailing and the leading sides of the arm the mean O/H residuals (Fig.~\ref{fig:D16_NEarm}a) decrease  by $\delta\log$(O/H)\,=\,0.08 (0.01) dex, virtually the same amount found in Section~\ref{NEarmanalysis}.

It is instructive to point out that the apparent O/H enhancement at $(s,d)\simeq(0.8, 0.3)$~kpc, also evident in Fig.~\ref{fig:D16map}, is likely spurious, and a consequence of the fact that here the N/O ratio appears to be enhanced (see Fig.~\ref{fig:NEarm}c). Since the \citet{Dopita:2016} abundance diagnostic is based on photoionization models that incorporate a prescribed trend of N/O as a function of O/H, and the $y$ parameter in Eq.~\ref{Eq:y} increases with the \nii\ line strength, a physical enhancement in the N/O ratio (which we tentatively attribute to Wolf-Rayet star winds) would translate into a larger O/H value using Eq.~\ref{Eq:oh}.


\bsp	
\label{lastpage}

\end{document}